\documentclass[prl,floatfix,superscriptaddress,twocolumn,aps,showpacs,amsmath,amssymb]{revtex4}
\usepackage[latin1]{inputenc}
\usepackage{graphicx}% Include figure files
\usepackage{epstopdf}

\newcommand{\be}{\begin{equation}}
\newcommand{\ba}{\begin{eqnarray}}
\newcommand{\ea}{\end{eqnarray}}
\newcommand{\ee}{\end{equation}}

\newcommand{\XXX}[1]{}

\newcommand{\bea}{\begin{eqnarray}}
\newcommand{\eea}{\end{eqnarray}}

\newcommand{\AffilH}{\affiliation{Institut f\"ur Theoretische Physik, Ruprecht--Karls--Universit\"at Heidelberg, \\ Philosophenweg 19, D--69120 Heidelberg, Germany}}
\newcommand{\AffilCEA}{\affiliation{Institut de Physique Th\'eorique, CEA, IPhT and CNRS, URA2306, Gif Sur Yvette, F--91191, France}}
\newcommand{\AffilUSC}{\affiliation{Department of Physics, University of Southern California, Los Angeles, CA 90089-0484, USA}}

\usepackage{color}
\definecolor{darkred}{rgb}{0.6,0,0}
\definecolor{red}{rgb}{1,0,0}
\newcommand{\CommentPS}[1]{\textcolor{darkred}{PS: {#1}}}
\newcommand{\CommentHS}[1]{{\bf HS: {#1}}}
\newcommand{\CommentEB}[1]{\textcolor{red}{EB: #1}}

\renewcommand{\CommentPS}[1]{}
\renewcommand{\CommentHS}[1]{}

\renewcommand{\CommentEB}[1]{}

\begin{document}
\title{Quantum fluctuation theorem in an interacting setup: point contacts in fractional quantum Hall edge state devices}
\author{A. Komnik}\AffilH
\author{H. Saleur}\AffilCEA\AffilUSC

\begin{abstract}
We verify the validity of the Cohen-Gallavotti fluctuation theorem   for the strongly correlated problem of charge transfer through an impurity in a chiral Luttinger liquid, which is realizable experimentally as a quantum point contact in a fractional quantum Hall edge state device. This is accomplished via the development of an analytical method to calculate the full counting statistics (FCS) of the problem in all the parameter regimes involving the temperature, the Hall voltage, and the gate voltage.
 \end{abstract}
\pacs{05.60.Gg, 02.30.Ik, 71.10.Pm, 05.30.-d}
\maketitle
%%
%%
%%%%%%%%%%%%%%%%%%%%%%%%%%%%%%%%%%%%%%%%%%%%%%%%%%%%%%
%%
%%%%%%%%%%%%%%%%%%%%%%%%%%%%%%%%%%%%%%%%%%%%%%%%%%%%%%
%%

Fluctuation theorems (FT), which  capture universal properties of systems far from equilibrium, have been the subject of intense activity in the past fifteen years, both in the classical and in the quantum case \cite{Esposito2009}. One of the most celebrated examples is the Cohen--Gallavotti FT (CGFT) \cite{Evans1993,Gallavotti1995} about the ratio of the probability of having a time averaged entropy production $\overline{s}_t$ in a steady, nonequilibrium state, take the value $s$ and the probability that it takes the opposite value $-s$. The CGFT says that this ratio is  a simple exponential:
\begin{equation}
{p(\overline{s}_t=s)\over p(\overline{s}_t=-s)} =e^{st}\label{CGFT} \, .
\end{equation}
Nonequilibrium fluctuations are most naturally observed in open quantum systems exchanging energy or matter with reservoirs. Nanoscale electronic devices, where fluctuations can be resolved at the single electron level, provide an ideal setup to investigate experimentally and theoretically the CGFT. It is indeed easy to relate (\ref{CGFT}) to a symmetry of the full  counting statistics (FCS) generating function. Imagine for instance electronic transport between two leads at the same (inverse) temperature $\beta$ but different chemical potentials, coupled by a certain tunneling hamiltonian.
The generating function of charge transferred (for more precise definition, see below) is then expected to obey, in the large time limit \cite{Esposito2009}
\begin{equation}
\chi(\kappa)=\chi(\beta V-\kappa)\label{CGFT1}
\end{equation}
where $V$ is the difference of potential, and we have set the electronic charge equal to unity. The verification that (\ref{CGFT1}) holds in the noninteracting case for the so-called Lesovik--Levitov  (LL) formula \cite{Levitov1996} is a corner stone of the rapidly developing field of electron counting statistics \cite{Roeck2007}. This formula reads
\begin{equation}
\ln\chi(\kappa)=t\int de\ n(e)\ln\left[1+\tau(e) {\cosh(\kappa-{V\over 2T})-\cosh{V\over 2T}\over \cosh {e\over T}+\cosh{V\over 2T}}\right]\label{hope}
\end{equation}
where $t$ is the waiting time, $n(e)$ is the density of states per unit length (or time, after setting the value of sound velocity to unity) at energy $e$, and $\tau(e)$ is the modulus square of a scattering amplitude associated with the tunneling process. Expansion in powers of $\kappa$ gives rise to highly non trivial sum rules between the cumulants of order $p$ and all cumulants of higher order.

It certainly seems desirable to investigate relation (\ref{CGFT1}) in the presence of interactions. Unfortunately, this requires, on the theoretical side,  tackling interacting, out of equilibrium, strongly fluctuating
open quantum systems, which is in general a daunting task even for classical systems \cite{Lazarescu2011}. We can nevertheless rely on a lot of recent progress. The idea of using the Bethe ansatz to describe transport through quantum impurities has been pursued in several works \cite{Fendley1995,Mehta2006}, leading, in some cases,  to the calculation of full $I-V$ characteristics \cite{Boulat2008}, the second (noise) \cite{Fendley1996} and third  \cite{Komnik2006} cumulants of the current, and $T=0$ full counting statistics  \cite{Saleur2001}. While the approach led to some controversies, it has been confirmed repeatedly by perturbative expansions \cite{Saleur2001,Honer2010} and thorough numerical t-DMRG calculations \cite{Boulat2008,Branschadel2010}.
In fact, a very recent paper reports verification of the formulas in \cite{Saleur2001} for the FCS at $T=0$ \cite{Carr2011} over the whole crossover region between weak and strong effective couplings.

Building on this progress, we report in this paper the calculation of FCS at non zero temperature in an interacting problem, together with the verification of the CG symmetry (\ref{CGFT1}).

The basic system we consider is the  1D chiral Luttinger liquid
(CLL) with an impurity, which  describes in particular tunnelling
between fractional quantum Hall (FQH) edges \cite{Wen1990}, or tunneling in the interacting resonant level model (IRLM) at the self-dual point \cite{Boulat2008}.

The bulk Hamiltonian  can be written in terms of right/left ($R$,$L$) moving charge current
densities,
\begin{eqnarray}
  H_0 = \frac{\pi}{g} \, \int_{-\infty}^\infty \, d x \, \left( j_L^2 + j_R^2
  \right) \label{hamil}\, ,
\end{eqnarray}
where we fix $h=1$ and where $g$ is the Luttinger liquid
interaction parameter \cite{Kane1992}. In the case of a FQH
device $g=\nu$ when $1/\nu$ is an odd integer \cite{Wen1990}.
The original fermions can be rewritten as exponents of the phase
fields $j_{L,R} = \mp
\partial_x \phi_{L,R}/2 \pi$, so that the (local at $x=0$)
backscattering term (or interedge tunnelling) is given by
\begin{eqnarray}
 H_{BS} = \lambda_{BS} \cos\left[ \phi_L(0) - \phi_R(0) \right]\label{scatter} \, ,
\end{eqnarray}
where $\lambda_{BS}$ is the respective amplitude.
It has been shown in \cite{Fendley1995}, that $H_0+H_{BS}$ can be mapped
(after a folding and a transformation into an even-odd basis)
onto the boundary sine-Gordon (BSG) model, which is integrable \cite{Ghoshal1994}.

The strategy to calculate properties in the steady state relies on a scattering description of the transport process. This is partly related to the  Landauer--B\"uttiker approach, although the latter was developed for free theories, while we are dealing here with a (strongly) interacting situation. The point is that there exists  a basis of  quasiparticles for the Hilbert space where:  $(a)$ the initial reservoir--like boundary conditions can be readily applied; and $(b)$ scattering through the impurity is simply described in a way that allows control of the charge flowing through the system. More precisely, this basis, at $1/\nu$
odd integer -- to which we restrict now
 -- is made of kinks, antikinks (denoted by
subscripts $\pm$), and
$j=1 \dots 1/\nu-2$ breathers. All of them are massless and have
the dispersion relations $e_\pm(\theta) = e^\theta/2 $ and $e_j(\theta)=\sin[ \pi
j/2(1/\nu - 1)]\, e^\theta$, where $\theta$ is referred to as
rapidity and parameterizes the particle momenta. While the
breathers are neutral, the (anti)kinks carry charge (in units of the
electron charge) $Q=\mp 1$, where by charge we mean the difference between the original  charges in the L and R channels of (\ref{hamil}).

Of course, these quasiparticles are very complicated objects in terms of the bare electrons, but since all one wants to follow is the distribution of the charge, this does not matter.
The point is that, because of integrability, the quasiparticles  scatter without particle production, and the process in the large waiting time limit is entirely encoded in a {\sl scattering matrix}, with all the one particle -- diagonal as well as off-diagonal  -- elements given in \cite{Fendley1995}. We define the FCS generating function in the way discussed for instance in \cite{Klich2003}, by directly using the charge $Q$ (instead of the currents). This  charge  enjoys both thermal and quantum fluctuations. We assume the system is initially prepared with decoupled channels at potential $\pm V/2$, and we are interested in the large time limit, where the system settles in a stationary process.
A key feature of the integrable quasiparticle basis is that the DC component of the current  acts diagonally on the multiparticle states,  and the contributions to the FCS of different  rapidities  in the stationary limit simply factorize \footnote{In the time dependent case, the current operator does not conserve the number of particles any longer, although of course it conserves the charge.}. The argument in \cite{Klich2003} can then be straightforwardly generalized at every rapidity. Like in the free fermion case, one obtains in the end an expression for the FCS as the average of a certain, $\kappa$ dependent expression calculated with respect to an equilibrium distribution in the reservoirs. Going through the mapping of the initial problem onto the BSG model one ends up with
\begin{eqnarray}            \label{ourFCS}
 \chi(\kappa) = \Big\langle \prod_e
 \Big\{ 1 + {\tau}(e) \Big[ (e^{\kappa} - 1) \eta_- (1-\eta_+)
 \nonumber \\
 + (e^{-\kappa} - 1)
  \eta_+ (1-\eta_-) \Big] \Big\} \Big\rangle \, ,
\end{eqnarray}
where the  $\eta_{\pm}$ are occupation numbers (equal to zero or
one) for  single particle kink and antikink states of momentum  (energy) $e$
in  the equilibrium (and unfolded) integrable model. The brackets denote average
over the equilibrium distribution determined by the
potential $V$, and $\tau(e)$ is the probability that a kink (antikink) goes through without being scattered into an antikink (kink).

Hidden in the apparent simplicity of Eq.~(\ref{ourFCS}) is a fundamental subtlety: the quasiparticles are not free, and we must think of them as a gas of particles whose wave functions satisfy a rapidity dependent statistics. This is the price to pay for the simple description of the tunneling physics in terms of single particle scattering. The properties of bulk wave functions depend on the bulk sine-Gordon scattering matrix, whose elements in the massless limit are also given in \cite{Fendley1995}. The average in Eq.~(\ref{ourFCS}) therefore can only  be done using the thermodynamic Bethe ansatz technique. This means that the different levels $e$'s in the product are not independent, but are solutions of a system of coupled Bethe ansatz equations. Like in the noninteracting systems one can
introduce the level densities (per unit length and rapidity) $n_i(\theta)$
and filling fractions
$f_i(\theta)$, $i=\pm, 1, \dots 1/\nu-2$, which can be combined to
give the
density of occupied states $P_i (\theta) = n_i(\theta)
f_i(\theta)$. But in contrast with the noninteracting case, the level density at rapidity $\theta$   depends on the filling fractions at all other rapidities.

Formula (\ref{ourFCS}) can immediately be used to recover known expressions for the current and the DC noise. In the former case for instance, expanding to first order in $\kappa$ gives readily
$ I = I(V,T_B) = \int \, d \theta \, {\cal \tau}(\theta)
\langle ({P}_+
 - P_-\rangle(\theta)$. Here $\langle P_\pm\rangle$ denotes the average densities of kinks and antikinks in equilibrium on the line (without scattering) at temperature $T$ and potential $V$. The scale $T_B$ is a characteristic energy scale for the tunneling process, formally analog to the Kondo temperature in the Kondo problem. Its dependence on the microscopic tunneling amplitude $\lambda_{BS}$ in Eq.~(\ref{scatter}) is not universal; by scaling one has  $T_B\propto \lambda_{BS}^{1/1-\nu}$.  The physics of this expression is well known, and describes the crossover between a regime where essentially no current is backscattered and $I\approx \nu V$ to a regime where most current is backscattered. While the known expression for the DC noise \cite{Fendley1996} can be recovered as well, Eq.~(\ref{ourFCS}) gave readily access to the third cumulant, which was analyzed in \cite{Komnik2006}.

The calculation of the full FCS from Eq.~(\ref{ourFCS}) is a challenging problem, and we will  only give the main ideas here, postponing details to further publication. The first step is to write the generic term in the product in Eq.~(\ref{ourFCS}) as an exponential using the simple identity
\begin{eqnarray}
1+\tau\left[(e^\kappa-1)\eta_-(1-\eta_+)+(e^{-\kappa}-1)\eta_+(1-\eta_-)\right]
\nonumber\\
=\exp\left[\alpha\eta_-(1-\eta_+)+\beta\eta_+(1-\eta_-)\right]
\end{eqnarray}
with $\alpha(\beta)=\ln\left[1+\tau(e^{\pm \kappa}-1)\right]$. The next step is to go to a continuum limit where we replace discrete sets of allowed momenta by continuously varying rapidities, with associated densities. Note that in this replacement, the effective length of the system is in fact the waiting time $t$ (where we have set $v_F=1$). While a naive replacement $\eta_\pm\rightarrow f_\pm$ was used in \cite{Komnik2006}, it is not sufficient beyond the third cumulant. The point is that, in a continuum description, with a certain number of allowed levels $n_\pm (\theta)d\theta$ for the $\pm$ species, and a certain fraction $f_\pm(\theta)$ of these levels occupied, the `overlap' $\eta_+(\theta_i)\eta_-(\theta_i)$ for the (still discrete) $\theta_i\in [\theta,\theta+d\theta]$ can still strongly fluctuate. It is thus necessary to perform the sum over all the values of this overlap first, which can be done, in the large waiting time limit $t \rightarrow \infty$, by a saddle point technique \footnote{We remark that this difficulty occurs as well in the free case, where, although the final result is known -- see Eq.~(\ref{hope}), obtaining it via TBA is already quite difficult.}.  Once this is done, we can transform the average into a functional integral over the densities \cite{Yang1969}, taking into account the usual TBA entropic  contributions. The end result is of the form
\begin{equation}                 \label{chiviaZ}
\chi= Z^{-1} \,\int {\cal D}[n_\pm,n_i;f_\pm,f_i] \, \exp\left(S-{E\over T}+S_\kappa\right) \, ,
\end{equation}
where $Z$ is the expression in the numerator without $S_\kappa$, the energy has the form $E=t\sum_i\int d\theta e_i(\theta)P_i(\theta)$ with $e_\pm(\theta)=e^\theta/2 \mp V/2$ (the breather energies being unaffected by the voltage) and the entropy $S=-t\sum_i\int d\theta n_i(\theta)\left[f_i\ln f_i+(1-f_i)\ln(1-f_i)\right](\theta)$. The last term is of the form
\begin{equation}
S_\kappa=t\int n(\theta)d\theta \left[\alpha f_-(1-f_+)+\beta f_+(1-f_-)+\Delta(f_+,f_-)\right]
\end{equation}
where $n(\theta)=n_\pm(\theta)$, and $\Delta$ takes into account the fluctuations of overlaps between $\pm$ species in Eq.~(\ref{ourFCS}). It is given by
\begin{eqnarray}
 \Delta(f_+,f_-) = (\alpha + \beta) f_- f_+ + \ln \left[ \frac{(1-f_+)(1-f_-)}{1 - f_+ - f_- + f} \right]
 \nonumber \\
 + \sum_{j=\pm} f_j \, \ln \left[ \frac{f_j ( 1 - f_+ - f_- + f)}{( 1 - f_j)(f_j - f)} \right] \, ,
\end{eqnarray}
where $f$ satisfies the equation
\begin{equation}
 e^{\alpha + \beta} = \frac{f ( 1 - f_+ - f_- + f)}{(f_+ - f)(f_- - f)}
\end{equation}
and represents the distribution of states populated by both kinks and antikinks.

Technically the evaluation of the functional integrals (\ref{chiviaZ}) is done in several stages.
The saddle point conditions with respect to filling fractions produce $1/\nu$ equations for the required $2/\nu$ variables $n_\pm,n_i$ and $P_\pm,P_i$. The Bethe ansatz equations (see e.~g. \cite{Fendley1995} for further details) then yield the relations between densities of states and densities of occupied states $P_j$ and are exactly the same as for the evaluation of conventional observables,
\begin{eqnarray}
 n_i = e_i(\theta) + \sum_j \int d \theta' \,
 \Phi_{ij}(\theta - \theta') \, P_j(\theta') \, ,
\end{eqnarray}
where $i,j$ refer to (anti)kinks as well as breathers and $\Phi_{ij}(\theta)$ are bulk scattering matrices given in \cite{Fendley1995}.
The densities  $n_\pm$ are identical.

We performed this program for the nontrivial case of filling fraction $\nu=1/3$, where there is only one breather degree of freedom. The saddle point equations have then the following form
\begin{eqnarray}
 X_1(\theta) &=& \int d \theta' \, \Phi_{11}(\theta - \theta') \, Y_1(\theta') + \Phi_{1+}(\theta - \theta') \, Y_0(\theta') \nonumber \\
 X_+(\theta) &=& \int d \theta' \, \Phi_{1+}(\theta - \theta') \, Y_1(\theta') + \Phi_{++}(\theta - \theta') \, Y_0(\theta') \nonumber \\
 X_-(\theta) &=& X_+(\theta) \, ,
\end{eqnarray}
where $Y_1 = \ln(1 - f_1)$, $X_1 = - e_1/T + \ln (f_1^{-1} - 1)$, and
\begin{widetext}
 \begin{eqnarray}
  X_\pm &=& \beta,\alpha - \frac{e_\pm \mp V/2}{T} + \ln \left( \frac{1 - f_+ - f_- + f}{f_\pm - f} \right) + \frac{f_\mp \, }{f_\mp - f}\frac{\partial f}{\partial f_\pm} + \frac{f ( f_\mp - 1)}{( 1 - f_- - f_+ + f)(f_\pm - f)} \left( 1 - \frac{\partial f}{\partial f_\pm} \right) \, , \\
  %%%%%%%
  \nonumber
  Y_0 &=& \ln ( 1- f_+ - f_- + f) + \frac{(1 - f_- - f_+)(f_+ + f_-)}{1 - f_+ - f_- + f} - \sum_{j=\pm} \frac{f_j^2}{f_j - f} + \left( \sum_{j=\pm} \frac{f_j}{f_j - f} - \frac{1 - f_+ - f_-}{1 - f_- - f_+ + f} \right) \sum_{j=\pm} f_j \frac{\partial f}{\partial f_j} \, ,
 \end{eqnarray}
\end{widetext}
and represent a closed equation system with respect to filling fractions $f_i$.
They are conveniently solved for numerically for a wide range of parameters. In the second step one evaluates the  $P_i$'s and calculates the cumulant generating function.

\begin{figure}
\vspace*{0.5cm}
\begin{center}
\includegraphics[scale=.35]{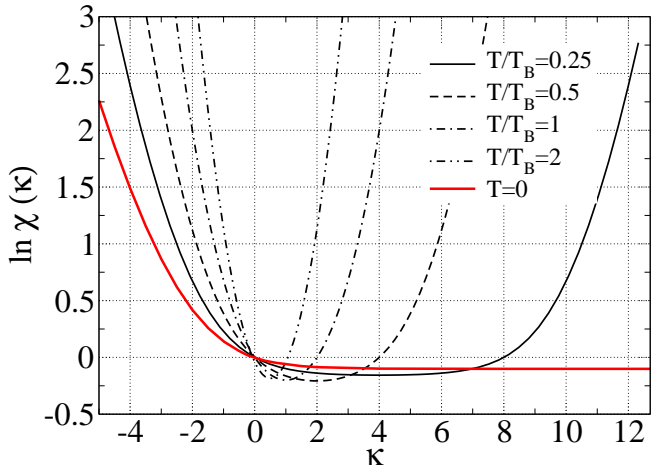}
\caption[]{\label{g13plotV2_different_T}(Color online) FCS generating function normalized to $2tT_B/h$ for $V/T_B=2$ and different temperatures.} \vspace*{-0.5cm}
\end{center}
\end{figure}

\begin{figure}
\vspace*{1.5cm}
\begin{center}
\includegraphics[scale=.35]{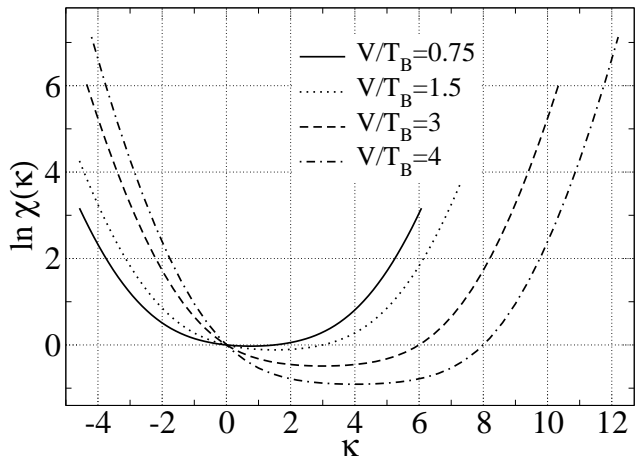}
\caption[]{\label{g13plotVforT05} FCS generating function normalized to $2tT_B/h$ at fixed temperature $T/T_B=0.5$ and different voltages. } \vspace*{-0.5cm}
\end{center}
\end{figure}

Similarly to the free case the FCS turns out to be a universal function of voltage $V/T_B$ and temperature $T/T_B$ measured in units of tunneling strength. With all fundamental constants restored (spinful case) one obtains
\begin{equation}
 \ln \chi(\kappa) = \frac{2 t}{h} T_B \, F(eV/T_B,T/T_B) \, .
\end{equation}
The function $F(V/T_B,T/T_B)$ is plotted in Figs.~\ref{g13plotV2_different_T} and \ref{g13plotVforT05} for different temperatures and voltages.
%%%%%%%%%%%%%%%%%%%%%%%%%%%%%%%%%%%%%%%%%%%%%%%%%%%%%%%%%%%%%%%%%%%%%
%%%%%%%%%%%%%%%%%%%%%%%%%%%%%%%%%%%%%%%%%%%%%%%%%%%%%%%%%%%%%%%%%%%%%
%%%%%%%%%%%%%%%%%%%%%%%%%%%%%%%%%%%%%%%%%%%%%%%%%%%%%%%%%%%%%%%%%%%%%
%%%%%%%%%%%%%%%%%%%%%%%%%%%%%%%%%%%%%%%%%%%%%%%%%%%%%%%%%%%%%%%%%%%%%
By virtue of normalization all curves go through the coordinate origin. However, all of them possess yet another zero point at precisely $\kappa = V/T=\beta V$, as is indeed prescribed by Eq.~(\ref{CGFT1}). Even more interesting is the overall curve symmetry with respect to reflections around the line $\kappa = V/2T$, which is again perfectly consistent with the above fluctuation theorem. While this is not surprising in the regimes of high temperatures and low voltages, when the system can be considered as being more `classical', it is most  interesting to verify  that this property persists at strong coupling as well. At $T=0$ the second zero moves to infinity and the curve approaches the `hockey stick' shape described by the analytical solution presented in \cite{Saleur2001}, see Fig.~\ref{g13plotV2_different_T}.

To conclude, we presented a method to determine the full  cumulant generating function of the nonequilibrium BSG model. It uses a modification of the thermodynamic Bethe ansatz and allows access to \emph{all} values of the counting field and system parameters. At finite temperature we confirm the validity of the CGFT in all regimes and  reproduce all known limiting cases. While the ultimate proof of the CGFT for generic interacting quantum system is still an open issue we have supplied an example of a nontrivial genuinely interacting system, in which it unmistakably holds. One avenue of future research might be the numerical evaluation of the FCS in the spirit of \cite{Carr2011}.

\acknowledgments AK is supported
by the Kompetenznetz ``Funktionelle Nanostrukturen III'' of the Baden-W\"urttemberg Stiftung, by the DFG under grant No.~KO~2235/3, and CQD and ``Enable fund'' of the University of Heidelberg. HS is supported by the ANR Projet 2010 Blanc SIMI 4 : DIME. We thank E. Boulat, K. Mallick and P. Schmitteckert for useful discussions.

%%
%%%%%%%%%%%%%%%%%%%%%%%%%%%%%%%%%%%%%%%%%%%%%%%%%%%%%%
%\bibliographystyle{plain}
\bibliography{CG_PRL}

\end{document}